\documentclass[trackchanges,twocolumn]{aastex7}

\begin{document}

\title{Observational study of chromospheric jets in and around a sunspot observed by NVST and SDO}

\author{Guotang Wu}
\affiliation{Yunnan Observatories, Chinese Academy of Sciences, Kunming 650216, People's Republic of China}
\affiliation{University of Chinese Academy of Sciences, Beijing 100049, People's Republic of China}
\email{wuguotang@ynao.ac.cn}

\author[0000-0003-2891-6267]{Xiaoli Yan}
\affiliation{Yunnan Observatories, Chinese Academy of Sciences, Kunming 650216, People's Republic of China}
\affiliation{Yunnan Key Laboratory of Solar Physics and Space Science, Kunming 650216, People's Republic of China}
\email{yanxl@ynao.ac.cn}

\correspondingauthor{Xiaoli Yan}
\email{yanxl@ynao.ac.cn}

\author[0000-0002-6526-5363]{Zhike Xue}
\affiliation{Yunnan Observatories, Chinese Academy of Sciences, Kunming 650216, People's Republic of China}
\affiliation{Yunnan Key Laboratory of Solar Physics and Space Science, Kunming 650216, People's Republic of China}
\email{zkxue@ynao.ac.cn}

\author[0000-0003-4393-9731]{Jincheng Wang}
\affiliation{Yunnan Observatories, Chinese Academy of Sciences, Kunming 650216, People's Republic of China}
\affiliation{Yunnan Key Laboratory of Solar Physics and Space Science, Kunming 650216, People's Republic of China}
\email{wangjincheng@ynao.ac.cn}

\author[0000-0002-9121-9686]{Zhe Xu}
\affiliation{Yunnan Observatories, Chinese Academy of Sciences, Kunming 650216, People's Republic of China}
\affiliation{Yunnan Key Laboratory of Solar Physics and Space Science, Kunming 650216, People's Republic of China}
\email{xuzhe6249@ynao.ac.cn}

\author[0000-0003-0236-2243]{Liheng Yang}
\affiliation{Yunnan Observatories, Chinese Academy of Sciences, Kunming 650216, People's Republic of China}
\affiliation{Yunnan Key Laboratory of Solar Physics and Space Science, Kunming 650216, People's Republic of China}
\email{yangliheng@ynao.ac.cn}

\author[0000-0002-0464-6760]{Yian Zhou}
\affiliation{Yunnan Observatories, Chinese Academy of Sciences, Kunming 650216, People's Republic of China}
\email{zhouyian@ynao.ac.cn}

\author{Liping Yang}
\affiliation{School of Physics, Electrical and Energy Engineering, Chuxiong Normal University, Chuxiong 675000, China}
\email{ylp@cxtc.edu.cn}

\author{Xinsheng Zhang}
\affiliation{Yunnan Observatories, Chinese Academy of Sciences, Kunming 650216, People's Republic of China}
\affiliation{University of Chinese Academy of Sciences, Beijing 100049, People's Republic of China}
\email{zhangxinsheng@ynao.ac.cn}

\author{Qifan Dong}
\affiliation{Yunnan Observatories, Chinese Academy of Sciences, Kunming 650216, People's Republic of China}
\affiliation{University of Chinese Academy of Sciences, Beijing 100049, People's Republic of China}
\email{dongqifan@ynao.ac.cn}

\author{Zongyin Wu}
\affiliation{Yunnan Observatories, Chinese Academy of Sciences, Kunming 650216, People's Republic of China}
\affiliation{University of Chinese Academy of Sciences, Beijing 100049, People's Republic of China}
\email{wuzongyin@ynao.ac.cn}

\begin{abstract}

To better understand the characteristics, driving mechanisms, and potential heating contributions of chromospheric jets, we analyze two contrasting types: one originating from within the sunspot penumbra (inside jets), and the other originating from outside the penumbra (outside jets). Statistical analysis of 100 jets (50 inside jets and 50 outside jets) reveals that inside jets have a projected velocity range of 4--14~km\,s$^{-1}$, a length range of 1--4~Mm, a width range of 0.2--0.6~Mm, and a lifetime range of 135--450~s, with mean values of 7.90~km\,s$^{-1}$, 2.61~Mm, 0.41~Mm, and 260~s, respectively. About 52\% of inside jets are associated with brightenings in H$\alpha$ blue wing images, and some show high-temperature signatures, suggesting a connection with localized energy release. In contrast, outside jets have higher velocities (8--50~km\,s$^{-1}$, average 19.04~km\,s$^{-1}$), greater lengths (average 6.26~Mm, up to 27.27~Mm), slightly larger widths (average 0.46~Mm), and longer lifetimes (135--630~s, average 327~s). They typically originate from regions of opposite magnetic polarities and are associated with magnetic flux emergence and EUV brightenings. Some outside jets correspond to coronal jets with inverted Y-shaped structures and temperatures exceeding one million Kelvin. Our results suggest that both jet types are driven by magnetic reconnection occurring in distinct magnetic field configurations and contribute to chromospheric and coronal heating.

\end{abstract}

\keywords{
Solar activity (\href{https://astrothesaurus.org/uat/1475}{1475}) --- 
Solar chromosphere (\href{https://astrothesaurus.org/uat/1479}{1479}) --- 
Solar magnetic fields (\href{https://astrothesaurus.org/uat/1503}{1503})
}

\section{Introduction} \label{sec:introduction}

Solar chromosphere is a highly dynamic layer above the photosphere, where energy is frequently released in the form of transient brightening events and jet-like ejections (see reviews by \citealp{2019ARA&A..57..189C} and \citealp{2021RSPSA.47700217S}). Recent observations have revealed that chromospheric jets occur in multiple regions, including active regions, quiet-Sun regions, and coronal holes (e.g., \citealp{2010ApJ...714L..31G,2011ApJ...728..103L,2023ApJ...945...96Y}). Since the plasma environments and magnetic field configurations vary across these locations, chromospheric jets exhibit different spatiotemporal characteristics. Advances in both high-resolution observations and numerical simulations have significantly enhanced our understanding of these chromospheric jets. However, their formation, evolution, and contribution to coronal heating and solar wind acceleration remain only partially understood. Therefore, a comprehensive study of chromospheric jets is essential for addressing key problems in solar physics.

The earliest observations of chromospheric jets in active regions, known as H$\alpha$ surges \citep{1973SoPh...32..139R}, revealed straight or slightly curved mass ejections associated with satellite sunspots or evolving magnetic structures \citep{1968IAUS...35...77R,1973SoPh...28...95R}. These surges typically ascend at speeds of 20 to 200~km\,s$^{-1}$, last for 10 to 20~min, and extend into the corona before falling back \citep{1973SoPh...28..477P,1994A&A...282..240G}. Several studies have suggested a connection between surges and Ellerman bombs, implying that the driving mechanism is magnetic reconnection \citep{2011ApJ...736...71W,2013ApJ...774...32V,2013SoPh..288...39Y}. In addition, \citet{1999ApJ...513L..75C}, \citet{2003PASJ...55..313Y}, and \citet{2004ApJ...610.1136L} presented compelling evidence that surges occur at sites of magnetic flux emergence and cancellation. This finding is consistent with the previously proposed mechanism of magnetic reconnection between the emerging flux and the pre-existing magnetic field \citep{1992PASJ...44L.173S,1993ASPC...46..507K,1995SoPh..156..245S,1996ApJ...464.1016C}. Multiwavelength observations further indicated that surges often coexist with EUV or X-ray jets, and that the cool components were delayed relative to their hot counterparts \citep{2007A&A...469..331J}. With the improvement of the spatial resolution of telescopes, a variety of chromospheric jets have been identified. These include spicules observed at the solar limb \citep{1968SoPh....3..367B,1972ARA&A..10...73B}, mottles seen on the solar disk \citep{1994A&A...290..285T}, as well as chromospheric anemone jets \citep{2007Sci...318.1591S} and dynamic fibrils \citep{2006ApJ...647L..73H} typically found in active regions, and even jets situated in sunspot light bridges \citep{2001ApJ...555L..65A,2018ApJ...854...92T}. Smaller in scale than the chromospheric jets mentioned above, penumbral microjets are tiny and transient jet-like features in the sunspot penumbra. Each event typically exhibits a width less than 400~km, a lifetime shorter than one minute, and an estimated energy on the order of nanoflares \citep{2007Sci...318.1594K}.

Chromospheric jets occurring in various solar regions, including regions in and around sunspot penumbra, exhibit diverse morphologies and characteristics depending on their locations and magnetic environments. Within the penumbra, penumbral microjets are small-scale jets aligned with penumbral filaments and exhibit more horizontal inclinations with increasing distance from the umbra \citep{2008A&A...488L..33J}. Subsequent studies employing automated detection methods have traced extensive samples of penumbral microjets to examine the spectral profiles in detail \citep{2017A&A...602A..80D}. Multiwavelength observations revealed signatures of penumbral microjets in transition region lines \citep{2015ApJ...811L..33V}, whereas no significant response was detected in coronal passbands \citep{2016ApJ...816...92T}. Outside the penumbra, chromospheric jets typically exhibit larger spatial scales and more dynamic behaviors, as demonstrated by events including H$\alpha$ surges and chromospheric anemone jets. The two-dimensional MHD simulation based on the emerging flux-reconnection model reproduced the structure and evolution of a giant jet observed by \citet{2008ApJ...683L..83N}. Transverse oscillations along the jet, interpreted as Alfvén waves, were consistently detected in both observations and simulations. However, the triggering mechanism of chromospheric jets in and around the sunspot remains elusive. This study offers new insights by conducting a comparative analysis of these two jet types, aiming to uncover their physical properties and formation processes.

In this paper, we investigate chromospheric jets located in and around a sunspot. The observations and data reduction methods are detailed in Section~\ref{sec:observations and data reduction}. Section~\ref{sec:results} presents the primary results of our analysis, and Section~\ref{sec:conclusions and discussion} offers the conclusions and discussion.

\section{Observations and data reduction} \label{sec:observations and data reduction}

\subsection{Observations} \label{subsec:observations}

Based on New Vacuum Solar Telescope (NVST) observations of active region NOAA 13386, we analyze two types of chromospheric jets: one originating from within the sunspot penumbra (hereafter referred to as inside jets), and the other originating from outside the penumbra (hereafter referred to as outside jets). The NVST is a 1-meter vacuum solar telescope located at the Fuxian Solar Observatory of Yunnan Observatories, Chinese Academy of Sciences \citep{2014RAA....14..705L,2020ScChE..63.1656Y}. It provides high-resolution images of the photosphere (TiO $7058\,\text{\AA}$) and chromosphere (H$\alpha$ $6563\,\text{\AA}$ and He I $10830\,\text{\AA}$). In this study, we primarily use the H$\alpha$ line center and blue wing ($-0.6$~\AA) images to investigate the physical properties of chromospheric jets. The field of view (FOV) of these H$\alpha$ images is $150^{\prime\prime} \times 150^{\prime\prime}$, with a cadence of 45~s and a spatial resolution of 0\farcs165 per pixel.

\begin{figure*}
  \centering
  \includegraphics[width=0.9\textwidth]{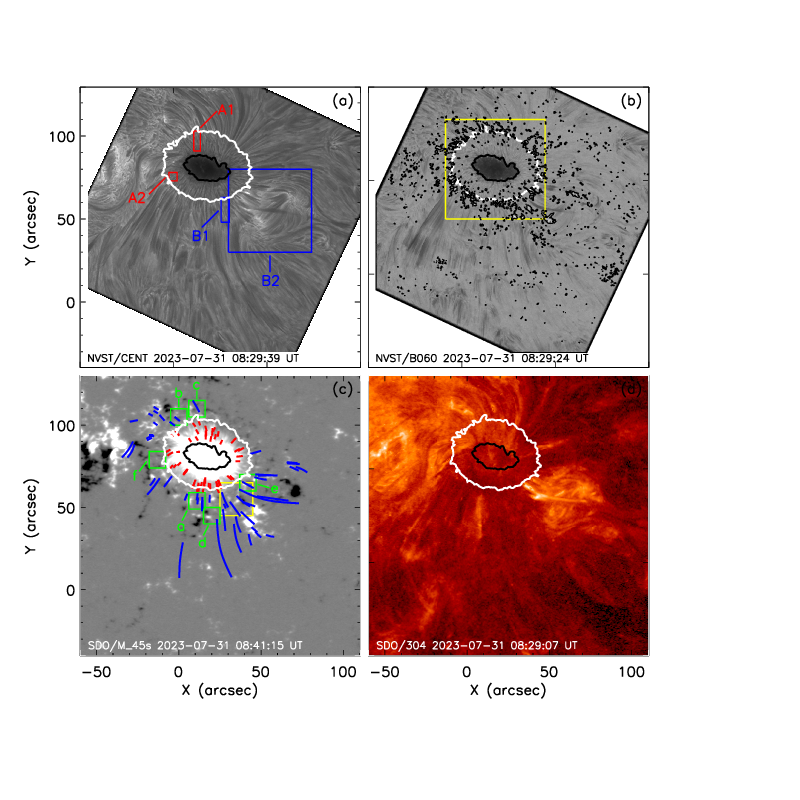}
  \caption{Overview of active region NOAA 13386. (a) H$\alpha$ line center image observed by NVST. The red and blue boxes outline subregions used in subsequent figures. (b) H$\alpha$ blue wing image at $-0.6\,\text{\AA}$. The black contours indicate brightenings where the intensity exceeds 140\% of the mean value over the field of view. The yellow box outlines the subregion shown in Figure~\ref{fig:figure4}. (c) Line-of-sight magnetogram from SDO/HMI, scaled from $-500$\,G to $500$\,G. Red lines represent 50 inside jets and blue lines represent 50 outside jets. The yellow and green boxes outline the subregions shown in Figures~\ref{fig:figure8} and \ref{fig:figure9}, respectively. (d) Corresponding SDO/AIA 304\,\AA\ image. The sunspot umbra and penumbra contours shown in panels~(a)--(d) are derived from the SDO/HMI continuum image, with the umbra marked in black and the penumbra in white.}
  \label{fig:figure1}
\end{figure*}

Figure~\ref{fig:figure1} shows snapshot images of the sunspot, which is situated at the disk center (N08, W01). Numerous chromospheric jets are widely distributed throughout the active region, as seen in the H$\alpha$ images in Figure~\ref{fig:figure1}(a) and~\ref{fig:figure1}(b). We divide the active region into two parts along the penumbral boundary, which is derived from the continuum image from the Helioseismic and Magnetic Imager (HMI; \citealp{2012SoPh..275..207S}) on board the Solar Dynamics Observatory (SDO; \citealp{2012SoPh..275....3P}). In Figure~\ref{fig:figure1}(c), which is the line-of-sight (LOS) magnetogram from SDO/HMI, the red lines represent 50 jets whose footpoints are clearly located in the penumbra, and blue lines indicate the other 50 jets originating from outside the penumbra. Additionally, partial simultaneous observations are obtained at seven EUV wavelengths from the Atmospheric Imaging Assembly (AIA; \citealp{2012SoPh..275...17L}) on board SDO, and we co-align the NVST H$\alpha$ image with the SDO/AIA $304\,\text{\AA}$ image as shown in Figure~\ref{fig:figure1}(d).

\subsection{Data reduction methods} \label{subsec:data reduction methods}

We perform visual inspections to identify those chromospheric jets that can be reliably tracked and subsequently measure their physical properties, such as projected velocity, length, width and lifetime. The velocity of a jet is derived from a time–distance diagram constructed by extracting a spatial slice along the jet axis from a series of H$\alpha$ blue wing images. The length is determined by the apparent maximum extent in H$\alpha$ line center images. In the same image, the width is defined as the full width at half maximum (FWHM) of the intensity profile, obtained by fitting the intensity profile perpendicular to the jet axis at a height free from contamination by other chromospheric material. The lifetime is measured from H$\alpha$ blue wing images, starting from the first frame in which the jet appears to the last frame in which its main structure is still identifiable. It is calculated as the number of frames multiplied by the temporal resolution of 45~s. Since our sample selection requires the jet to be visible in at least three consecutive frames, the derived lifetime is no less than 135~s.

The H$\alpha$ blue wing images are used to identify brightenings. As shown in Figure~\ref{fig:figure1}(b), we use 140\% of the mean intensity over the field of view as the threshold to contour bright features. The brightenings are predominantly concentrated around the sunspot, while a few are also found within the penumbra. When analyzing the outside jets in Figures~\ref{fig:figure8} and~\ref{fig:figure9}, we deliberately increase the threshold to 150\% of the mean intensity to better isolate the relevant brightenings.

The thermal properties of chromospheric jets are investigated using the differential emission measure (DEM) method, following the approach developed by \citet{2015ApJ...807..143C}. This method utilizes six EUV passbands from the SDO/AIA ($171\,\text{\AA}$, $193\,\text{\AA}$, $211\,\text{\AA}$, $131\,\text{\AA}$, $335\,\text{\AA}$, $94\,\text{\AA}$, ranked in order of response temperature from low to high). We compute the emission measure (EM) by integrating the DEM over a restricted temperature range of $\log T = 6.0$--$6.3$, which corresponds to plasma temperatures exceeding one million Kelvin. The DEM analysis enables the identification and quantification of coronal plasma components associated with chromospheric jets.

\section{Results} \label{sec:results}

\subsection{Properties of inside jets} \label{subsec:properties of inside jets}

\begin{figure*}
  \centering
  \includegraphics[width=0.9\textwidth]{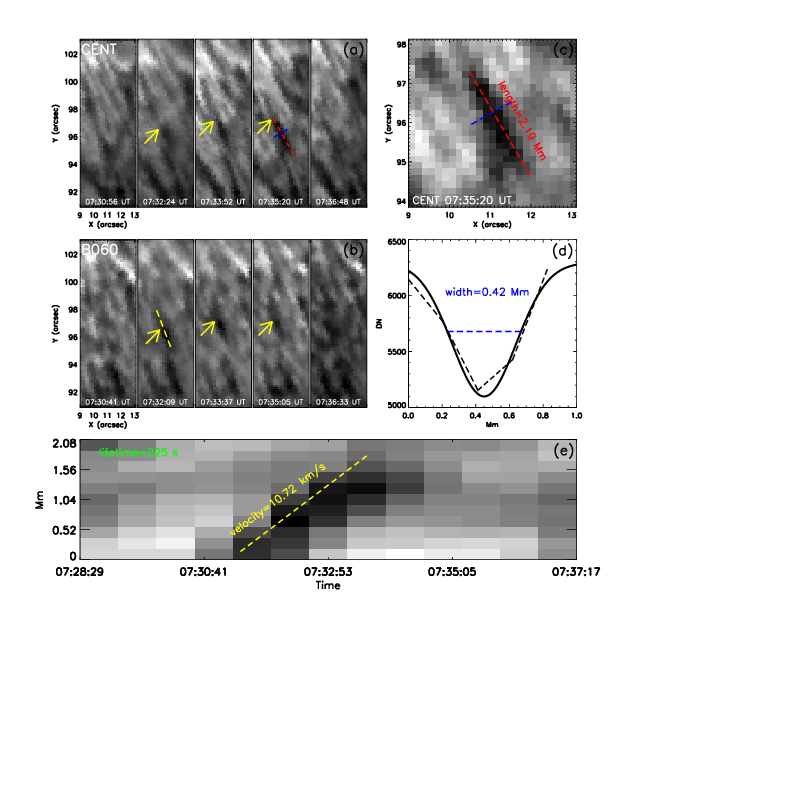}
  \caption{Evolution of an inside jet. (a) Time sequence of H$\alpha$ line center images corresponding to the box A1 in Figure~\ref{fig:figure1}(a). (b) Time sequence of H$\alpha$ blue wing images at $-0.6\,\text{\AA}$. Arrows in panels~(a) and (b) indicate the top of the inside jet. (c) Enlarged view of the fourth image of panel~(a). The red dashed line indicates the length of the inside jet, while the blue dashed line marks the position used to measure the width via intensity fitting. (d) Intensity profile along the blue dashed line in panel~(c), with the dashed curve showing the observed data and the solid curve depicting the Gaussian profile. (e) Time-distance diagram obtained along the yellow dashed line in the second image of panel~(b).  The yellow dashed line in panel~(e) is used to estimate the projected velocity, and the green annotation indicates the lifetime of the inside jet.}
  \label{fig:figure2}
\end{figure*}

Figure~\ref{fig:figure2} presents a typical example of an inside jet, which appears as a dark and collimated feature in both the H$\alpha$ line center and blue wing images. The footpoint of the inside jet is located within the penumbra, and its direction is aligned with that of the penumbral filaments. The physical properties of the inside jet are measured using the definitions and methods described in Section~\ref{subsec:data reduction methods}. The length is measured at 07:35:20~UT when it reaches its apparent maximum extent, along the red dashed line in Figure~\ref{fig:figure2}(c), with a value of approximately 2.19~Mm. The width is derived to be 0.42~Mm (FWHM) at the same time by fitting the intensity profile perpendicular to the inside jet axis along the blue dashed line. The intensity profile exhibits an inverted Gaussian shape, reflecting the absorption nature of the dark feature, as shown in Figure~\ref{fig:figure2}(d). Additionally, a time-distance diagram constructed along the inside jet axis indicates a velocity of about 10.72~km\,s$^{-1}$, as traced by the yellow dashed line in Figure~\ref{fig:figure2}(e). The inside jet has a lifetime of approximately 225~s.

\begin{figure*}
  \centering
  \includegraphics[width=0.9\textwidth]{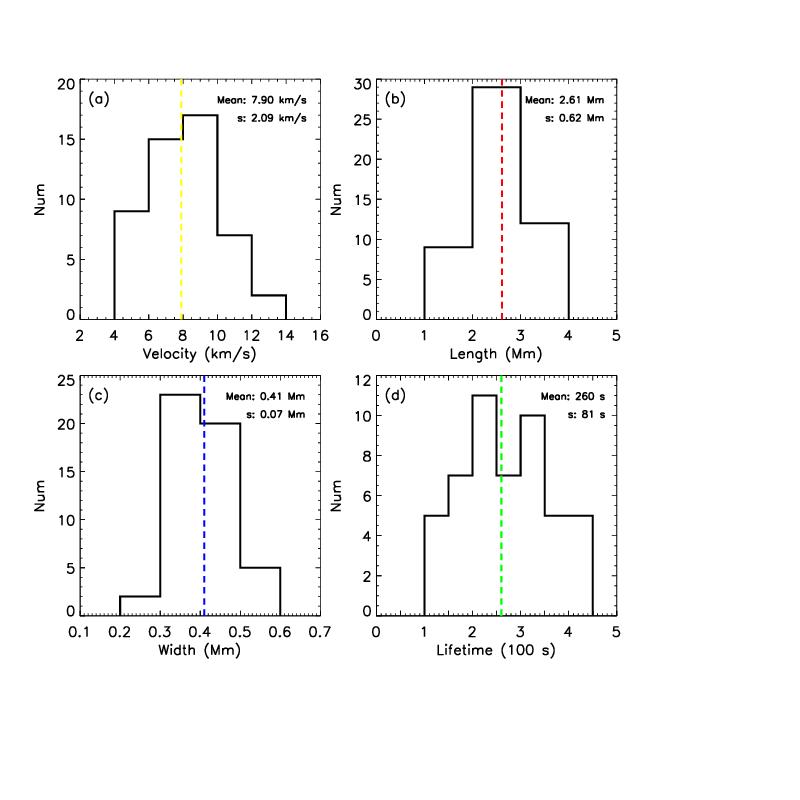}
  \caption{Statistical distributions of the physical properties of 50 inside jets. (a) Velocity distribution in km\,s$^{-1}$. (b) Length distribution in Mm. (c) Width distribution in Mm. (d) Lifetime distribution in 100~s. The yellow, red, blue, and green dashed lines in panels~(a), (b), (c), and (d) indicate the mean values of the velocity, length, width, and lifetime, respectively. The mean value and sample standard deviation are annotated in each panel.}
  \label{fig:figure3}
\end{figure*}

Following the methodology described above, the physical properties of 50 inside jets originating from within the penumbra are statistically analyzed. The histograms of the results are shown in Figure~\ref{fig:figure3}. The velocities of inside jets range from 4--14~km\,s$^{-1}$, with a mean value of 7.90~km\,s$^{-1}$. Their lengths span 1--4~Mm, averaging 2.61~Mm, while their widths range from 0.2--0.6~Mm, with an average of 0.41~Mm. The lifetimes range from 135--450~s, with a mean value of 260~s. The derived parameters are consistent with those of penumbral microjets observed in the H$\alpha$ blue wing, as reported by \citet{2019ApJ...876...47B}, but with longer lifetimes. All four distributions are nearly symmetric and peak close to their respective mean values, indicating that the physical properties of inside jets are characterized by well-defined typical scales with relatively small sample standard deviations. It should be noted that due to the limitations of spatial and temporal resolution, smaller and shorter-lived events may not be detected or fully resolved.

\begin{figure*}
  \centering
  \includegraphics[width=0.9\textwidth]{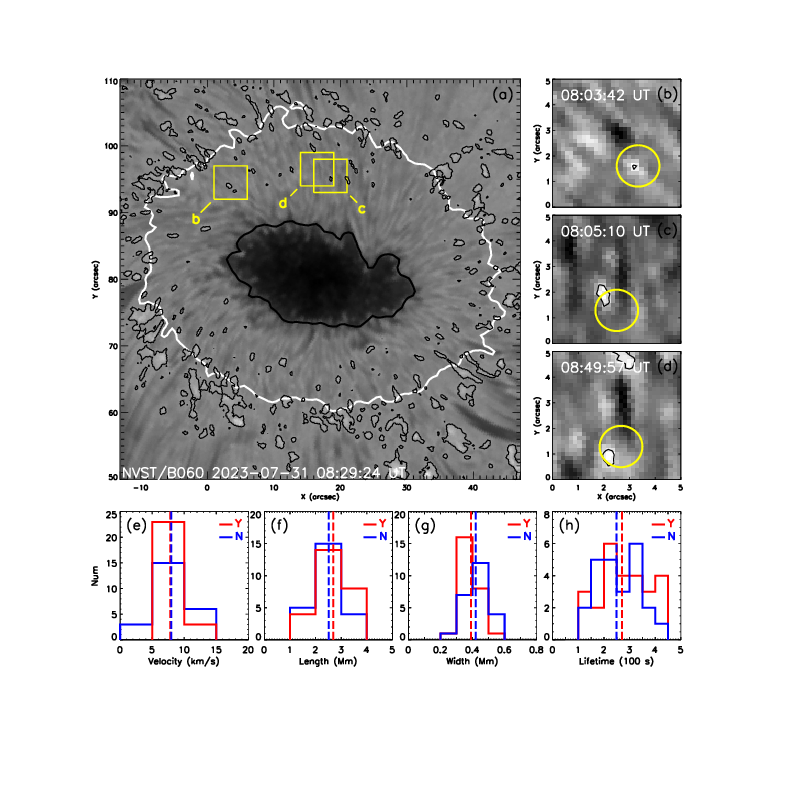}
  \caption{Brightenings identified in the H$\alpha$ blue wing at $-0.6\,\text{\AA}$ using a threshold of 140\% of the mean intensity. (a) Enlarged view of the sunspot corresponding to the yellow box in Figure~\ref{fig:figure1}(b). Three yellow boxes outline the subregions shown in panels~(b)--(d). (b)--(d): Three examples of inside jets associated with brightenings. Yellow circles highlight the footpoints of the jets. (e)--(h): Histograms comparing the physical properties of inside jets associated with brightenings (red, 26 samples) and without brightenings (blue, 24 samples).}
  \label{fig:figure4}
\end{figure*}

The field of view in Figure~\ref{fig:figure4} corresponds to the yellow box in Figure~\ref{fig:figure1}(b). The black contours indicate brightenings using a threshold of 140\% of the mean intensity. It is evident that the brightenings located within the penumbra are generally smaller and more sparsely distributed compared to those outside the penumbra. We further investigate whether the footpoints of inside jets are spatially related to these brightenings during their ejection. The right column displays three examples of inside jets associated with brightenings. In each case, a brightening appears in the vicinity of the jet footpoint. Among the 50 inside jets analyzed in our samples, 26 (52\%) are found to be associated with brightenings, while the remaining events show no clear spatial correlation. In the bottom row of Figure~\ref{fig:figure4}, we present histograms comparing the properties of inside jets associated with brightenings and those without such association. The histograms show no significant differences between the two groups in terms of velocity, length, width and lifetime.

\begin{figure*}
  \centering
  \includegraphics[width=0.9\textwidth]{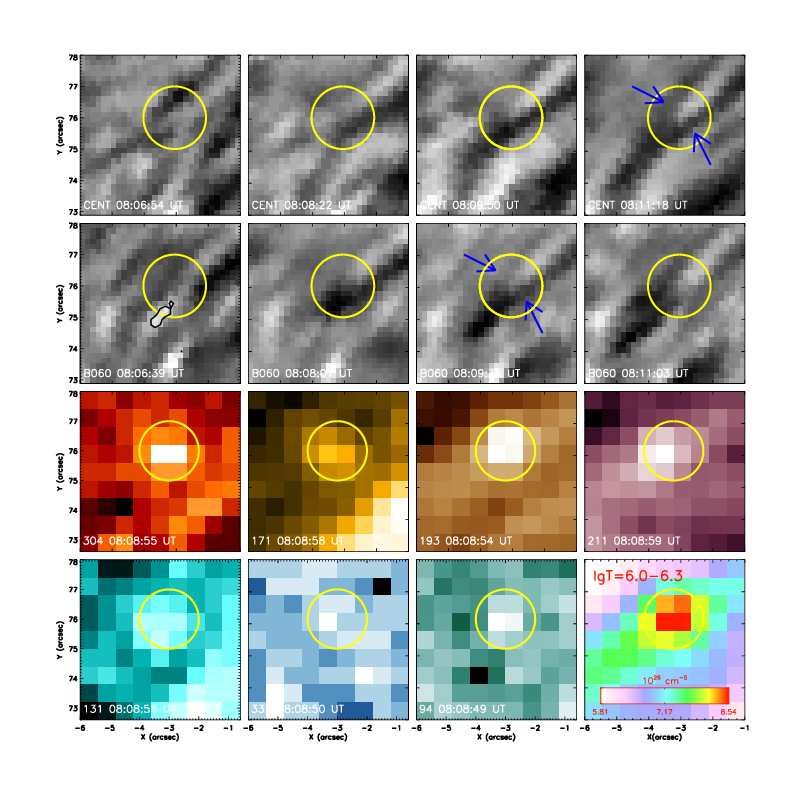}
  \caption{Multiwavelength observations of an inside jet corresponding to the box A2 in Figure~\ref{fig:figure1}(a). The first row shows the H$\alpha$ line center images observed by NVST. The second row displays the corresponding H$\alpha$ blue wing images at $-0.6\,\text{\AA}$. The black contours in the first panel of the second row indicate brightenings identified using a threshold of 140\% of the mean intensity. The third and fourth rows present observations from seven AIA EUV channels, arranged in order of increasing response temperature: $304\,\text{\AA}$, $171\,\text{\AA}$, $193\,\text{\AA}$, $211\,\text{\AA}$, $131\,\text{\AA}$, $335\,\text{\AA}$, and $94\,\text{\AA}$. The final panel shows the EM map integrated over the temperature range $\log T = 6.0$--$6.3$, derived from DEM analysis. Yellow circles highlight the footpoint of the inside jet, and blue arrows in some panels point to the two associated filaments.}
  \label{fig:figure5}
\end{figure*}

Figure~\ref{fig:figure5} shows the evolution of an inside jet. The inside jet originates from the intersection of two filaments, as indicated by the blue arrow in some panels. During the pre-ejection phase, brightenings are observed in the H$\alpha$ blue wing at 08:06:39~UT, which are indicated by the black contours. Furthermore, notable brightenings are detected in the AIA EUV channels when the inside jet is triggered. We perform the DEM analysis and find that the region has strong emission measures at $\log T = 6.0$--$6.3$. These observational features suggest that magnetic reconnection between two intersecting filaments may serve as the driving mechanism for inside jets. This process is consistent with the scenario proposed by \citet{2007Sci...318.1594K}, in which magnetic reconnection occurs between more horizontal and more vertical magnetic fields within the penumbra.

\subsection{Properties of outside jets} \label{subsec:properties of outside jets}

\begin{figure*}
  \centering
  \includegraphics[width=0.9\textwidth]{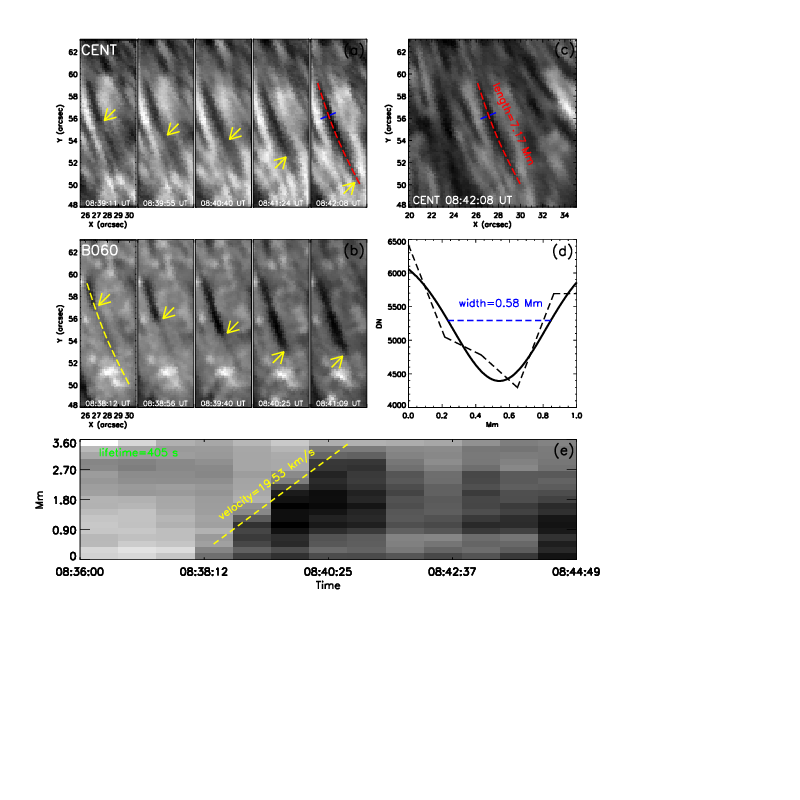}
  \caption{Evolution of an outside jet. (a) Time sequence of H$\alpha$ line center images corresponding to the box B1 in Figure~\ref{fig:figure1}(a). (b) Time sequence of H$\alpha$ blue wing images at $-0.6\,\text{\AA}$. Arrows in panels~(a) and (b) indicate the top of the outside jet. (c) Enlarged view of the fifth image of panel~(a). The red dashed line indicates the length of the outside jet, while the blue dashed line marks the position used to measure the width via intensity fitting. (d) Intensity profile along the blue dashed line in panel~(c), with the dashed curve showing the observed data and the solid curve depicting the Gaussian profile. (e) Time-distance diagram obtained along the yellow dashed line in the first image of panel~(b). The yellow dashed line in panel~(e) is used to estimate the projected velocity, and the green annotation indicates the lifetime of the outside jet.}
  \label{fig:figure6}
\end{figure*}

Figure~\ref{fig:figure6} shows a representative case of an outside jet. Similar to the example in Figure~\ref{fig:figure2}, the outside jet manifests as a collimated plasma ejection, but with a larger spatial scale. The same methods are applied to this outside jet, yielding a velocity of 19.53~km\,s$^{-1}$, a length of 7.17~Mm, a width of 0.58~Mm, and a lifetime of 405~s. In the H$\alpha$ images, the outside jet exhibits morphologies and characteristics similar to those of the inside jet, with the main differences being its greater velocity, length, width, and lifetime.

\begin{figure*}
  \centering
  \includegraphics[width=0.9\textwidth]{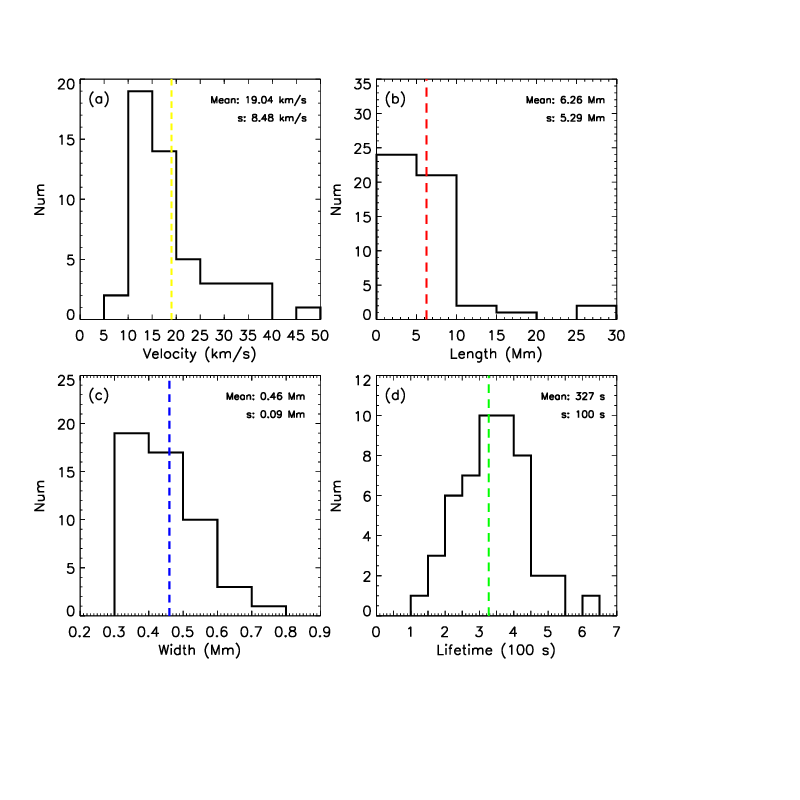}
  \caption{Statistical distributions of the physical properties of 50 outside jets. (a) Velocity distribution in km\,s$^{-1}$. (b) Length distribution in Mm. (c) Width distribution in Mm. (d) Lifetime distribution in 100~s. The yellow, red, blue, and green dashed lines in panels~(a), (b), (c), and (d) indicate the mean values of the velocity, length, width, and lifetime, respectively. The mean value and sample standard deviation are annotated in each panel.}
  \label{fig:figure7}
\end{figure*}

We also select 50 outside jets that can be reliably tracked and analyze their properties. The statistical results are presented in the histograms of Figure~\ref{fig:figure7}. The velocities of these outside jets mainly fall within the range of 8--50~km\,s$^{-1}$, with a mean value of 19.04~km\,s$^{-1}$, although the maximum reaches up to 49.30~km\,s$^{-1}$. The lengths are primarily distributed in the range of 1--20~Mm, with an average length of 6.26~Mm and a maximum of 27.27~Mm. Compared to inside jets, the widths of outside jets are slightly larger, with an average of 0.46~Mm. The lifetimes range from 135--630~s, with a mean value of 327~s. The physical properties of the outside jets closely resemble those of chromospheric anemone jets described by \citet{2011ApJ...731...43N}. Overall, the distributions of velocity, length, and width show a decreasing trend with increasing values, while the distribution of lifetime peaks at 300--400~s. The relatively large sample standard deviations indicate broader spreads of these properties.

\begin{figure*}
  \centering
  \includegraphics[width=0.9\textwidth]{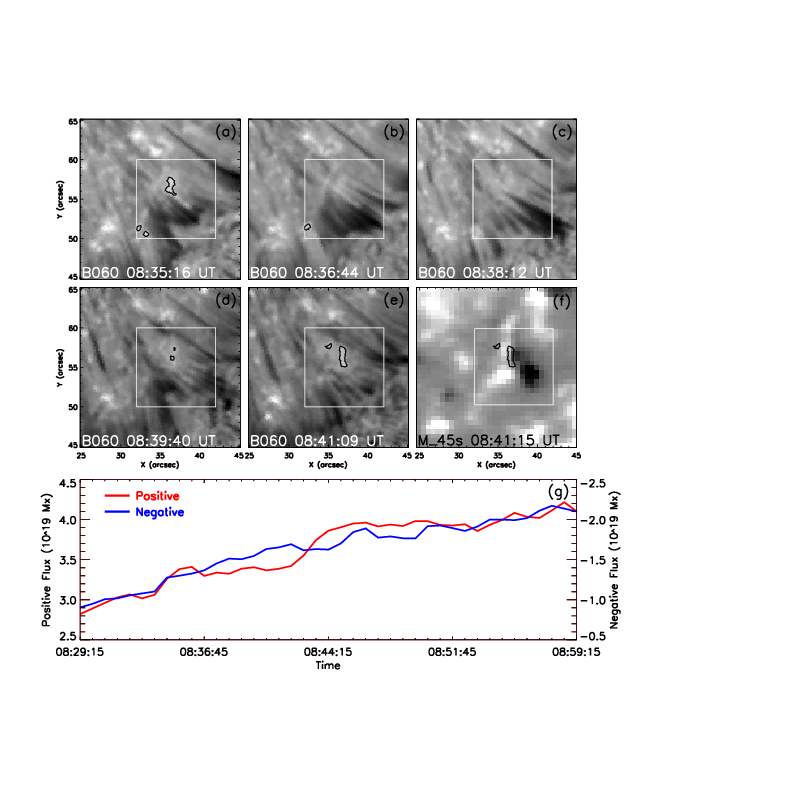}
  \caption{Magnetic activity of outside jets. (a)--(e) Time sequence of H$\alpha$ blue wing images at $-0.6\,\text{\AA}$ corresponding to the yellow box in Figure~\ref{fig:figure1}(c). The black contours indicate brightenings where the intensity exceeds 150\% of the mean value. (f) Line-of-sight magnetogram from SDO/HMI, scaled from $-100$\,G to $100$\,G. The black contours from panel~(e) are overplotted on the magnetogram. (g) Temporal evolution of the magnetic flux within the white box. Red and blue curves represent the positive and negative magnetic fluxes, respectively.}
  \label{fig:figure8}
\end{figure*}

\begin{figure*}
  \centering
  \includegraphics[width=0.9\textwidth]{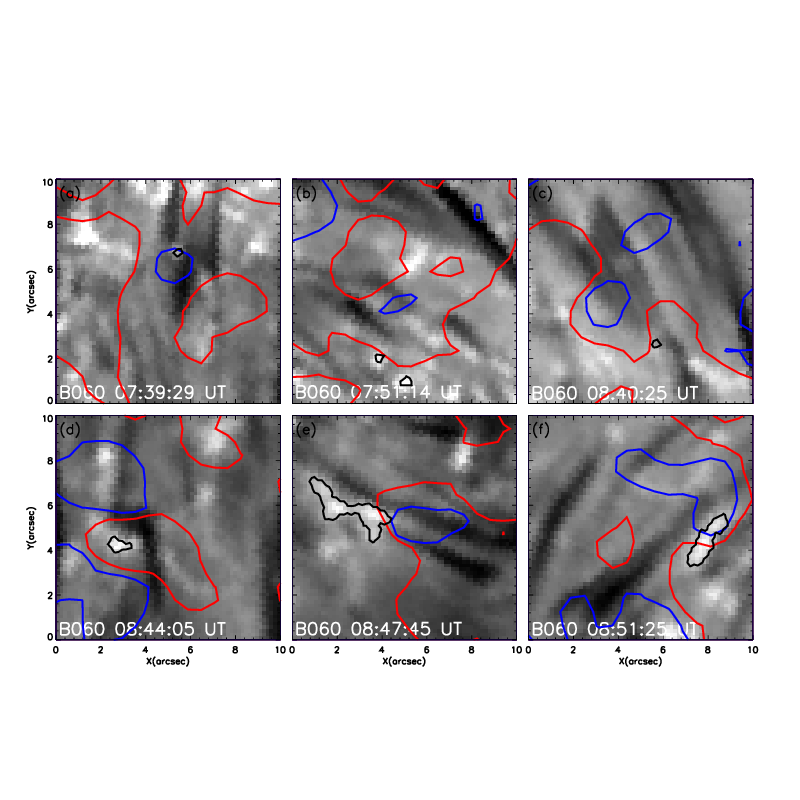}
  \caption{Examples of outside jets in H$\alpha$ blue wing images at $-0.6\,\text{\AA}$ corresponding to the green boxes in Figure~\ref{fig:figure1}(c). (a)--(c) Examples of outside jets that are included in the statistical sample. (d)--(f) Examples of outside jets that are excluded from the statistical sample. Red and blue contours represent the line-of-sight magnetic field strength at +50~G and -50~G, respectively. The black contours indicate brightenings where the intensity exceeds 150\% of the mean value, detected during the lifetimes of the outside jets.}
  \label{fig:figure9}
\end{figure*}

For outside jets located outside the penumbra, we can perform a detailed analysis of the magnetic field at their footpoints using line-of-sight magnetograms. The field of view in Figure~\ref{fig:figure8} corresponds to the yellow box in Figure~\ref{fig:figure1}(c), which is located outside the penumbra. Panels~(a)-(e) of Figure~\ref{fig:figure8} show the jet activities in this region. The brightenings are indicated by black contours using a threshold of 150\% of the mean intensity. From the magnetogram in panel~(f), the footpoint region is characterized by opposite magnetic polarities. The brightenings in the H$\alpha$ blue wing are located precisely at this position. We calculate the magnetic flux within the white box and present its temporal evolution in the bottom panel of Figure~\ref{fig:figure8}. The results indicate that both the positive and negative magnetic fluxes increased during this period. The emergence of magnetic flux may be responsible for the generation of the outside jets.

In addition, we examine the magnetic environments at the footpoints of more outside jets. As shown in Figure~\ref{fig:figure9}, we provide three examples from the statistical sample in panels~(a)--(c) and three cases excluded from the statistics in panels~(d)--(f). The latter were excluded owing to limited visibility or tracking uncertainty, yet all examples can be clearly identified as originating from regions of opposite magnetic polarities and are consistently associated with brightenings. We speculate that their formation is related to magnetic flux emergence and cancellation. This supports the three-dimensional simulations by \citet{2003Natur.425..692S} that demonstrate how magnetic flux emerging from the solar interior interacts with the ambient magnetic field.

\begin{figure*}
  \centering
  \includegraphics[width=0.9\textwidth]{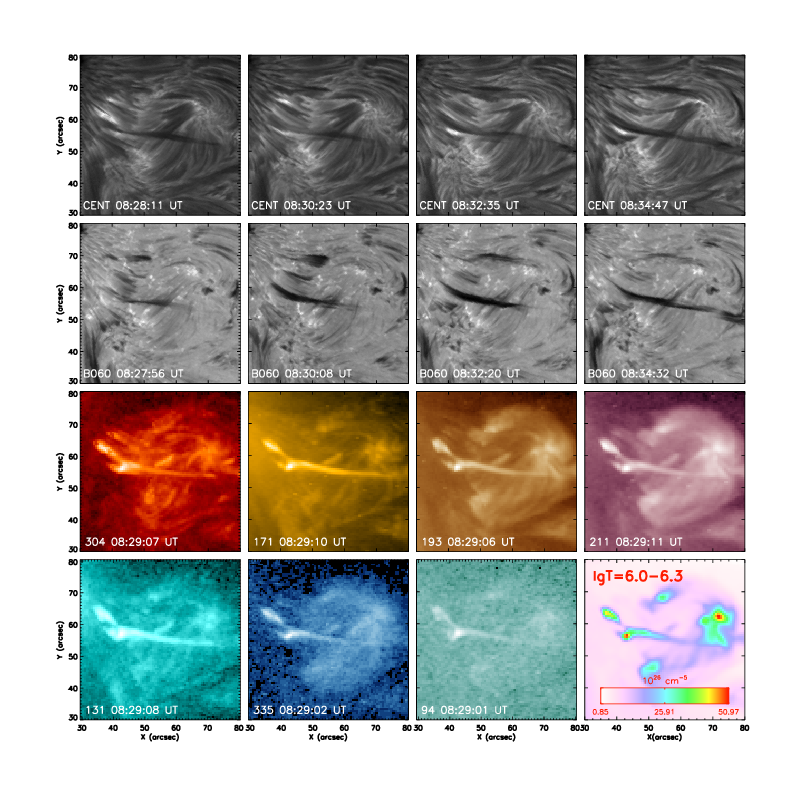}
  \caption{Multiwavelength observations of an outside jet corresponding to the box B2 in Figure~\ref{fig:figure1}(a). The first row shows the H$\alpha$ line center images observed by NVST. The second row displays the corresponding H$\alpha$ blue wing images at $-0.6\,\text{\AA}$. The third and fourth rows present observations from seven AIA EUV channels, arranged in order of increasing response temperature: $304\,\text{\AA}$, $171\,\text{\AA}$, $193\,\text{\AA}$, $211\,\text{\AA}$, $131\,\text{\AA}$, $335\,\text{\AA}$, and $94\,\text{\AA}$. The final panel shows the EM map integrated over the temperature range $\log T = 6.0$--$6.3$, derived from DEM analysis.}
  \label{fig:figure10}
\end{figure*}

Subsequently, a more prominent response is expected in high-temperature EUV channels compared to that observed in inside jets. As shown in Figure~\ref{fig:figure10}, a corresponding coronal jet can be identified in the AIA EUV channels for this outside jet. The coronal counterpart appears as an inverted Y-shaped structure with evident brightening at its footpoint. Based on these coronal observations and the results from DEM analysis, the coronal jet is inferred to be the hot component, generated by magnetic reconnection occurring in the upper chromosphere or lower corona, with plasma temperatures reaching up to one million Kelvin. The outside jet seen in the H$\alpha$ images corresponds to the cool component of the same reconnection-driven process. A similar relationship between surges and EUV jets has been identified by \citet{2007A&A...469..331J}.

\subsection{Comparison between inside and outside jets} \label{subsec:comparison between inside and outside jets}

\begin{figure*}
  \centering
  \includegraphics[width=\textwidth]{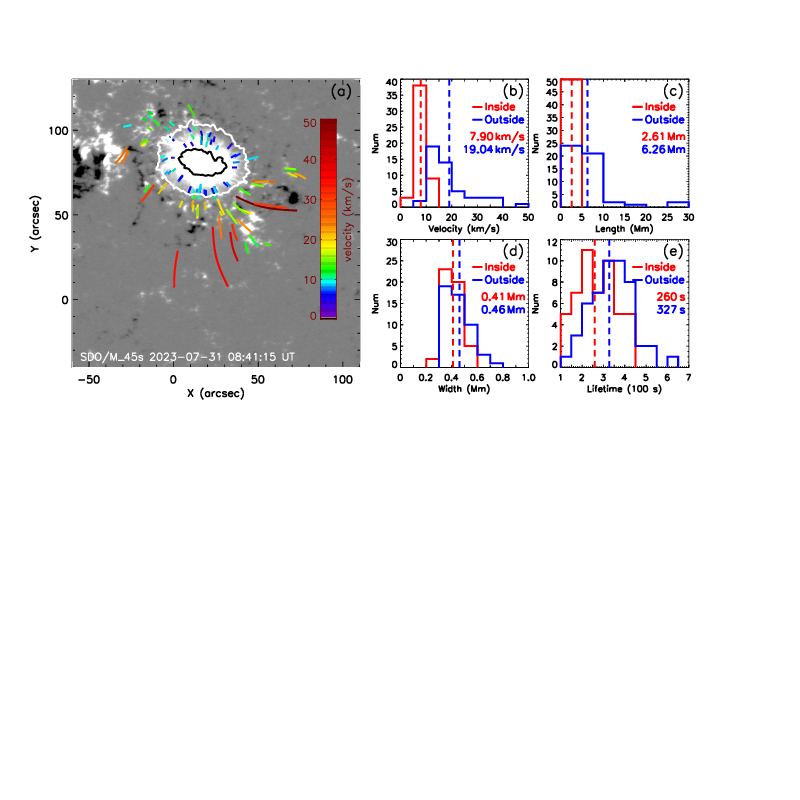}
      \caption{Contrast between inside and outside jets. (a) Line-of-sight magnetogram overlaid with 100 identified jets, with line colors indicating the projected velocities. (b)-(e) Histograms comparing the physical properties of two jet types. Inside jets are shown in red, while outside jets are shown in blue. The dashed lines indicate the mean values for each group, which are annotated in each panel.}
  \label{fig:figure11}
\end{figure*}

Inside and outside jets exhibit similar morphological characteristics, both appearing as narrow, collimated plasma ejections in the H$\alpha$ line center and blue wing images. These similarities suggest that the two types of jets may arise from a common physical mechanism operating under different plasma and magnetic environments. To visualize the spatial distribution and physical properties, the 100 jets are overlaid on a magnetogram in Figure~\ref{fig:figure11}(a). The line colors indicate the projected velocities, allowing for a clear comparison between the kinematic characteristics of inside and outside jets. The panel shows that chromospheric jets are distributed around the sunspot, either within the penumbra or in the surrounding regions. Moreover, outside jets tend to have higher velocities and exhibit greater lengths, widths, and lifetimes than inside jets.

A quantitative comparison further highlights the differences between inside and outside jets in Figure~\ref{fig:figure11} (b)--(e). Inside jets have lower velocities (4--14~km\,s$^{-1}$, mean 7.90~km\,s$^{-1}$), shorter lengths (1--4~Mm, mean 2.61~Mm), narrower widths (0.2--0.6~Mm, mean 0.41~Mm), and shorter lifetimes (135--450~s, mean 260~s). In contrast, outside jets exhibit significantly higher velocities (8--50~km\,s$^{-1}$, average 19.04~km\,s$^{-1}$), greater lengths (average 6.26~Mm, up to 27.27~Mm), slightly larger widths (average 0.46~Mm), and longer lifetimes (135--630~s, average 327~s). These differences are visually represented in the histograms, where inside jets exhibit more concentrated distributions around their mean values, whereas outside jets display broader and more dispersed distributions.

Despite differences in speed and size, both types of jets are likely driven by magnetic reconnection occurring in distinct magnetic field configurations. Inside jets typically originate within the penumbra, where component reconnection between intersecting filaments appears to trigger their formation. Outside jets are more commonly found in regions of opposite magnetic polarities, with their generation possibly related to magnetic flux emergence and cancellation. The consistent presence of brightenings in inside jets and magnetic activity in outside jets at their footpoints support the interpretation that magnetic reconnection is the common underlying mechanism.

In addition to their thermal properties, both inside and outside jets exhibit signatures in high-temperature EUV channels, indicating plasma heating to coronal temperatures. Notably, some outside jets display coronal counterparts in AIA observations, featuring inverted Y-shaped structures and brightenings at their footpoints that are hallmarks of magnetic reconnection events. The DEM analysis confirms that plasma temperatures during these events exceed one million Kelvin. These findings suggest that both small-scale inside jets and large-scale outside jets contribute to atmospheric heating.

\section{Conclusions and discussion} \label{sec:conclusions and discussion}

Based on NVST and SDO observations, we investigate two types of chromospheric jets. Both inside and outside jets exhibit collimated, jet-like structures in the H$\alpha$ line center and blue wing images. The two jet types have similar morphological characteristics, differing only in their spatial and temporal scales. On average, inside jets have lower projected velocities (7.90~km\,s$^{-1}$), shorter lengths (2.61~Mm), and narrower widths (0.41~Mm), and shorter lifetimes (260~s), whereas outside jets have higher projected velocities (19.04~km\,s$^{-1}$), greater lengths (6.26~Mm), and slightly larger widths (0.46~Mm), and longer lifetimes (327~s). These differences may be partly affected by projection effects. Inside jets, which likely originate within the penumbra, appear less inclined to the line of sight than outside jets in our viewing geometry. According to \citet{2008A&A...488L..33J}, the inclination of penumbral microjets increases with distance from umbra, ranging from about 35$^\circ$ at the umbra/penumbra boundary to 70$^\circ$ at the penumbra/quiet-Sun boundary. Even after correcting for projection effects using the inferred inclinations, the mean velocity and length of inside jets remain smaller than those of outside jets.

We examine brightenings observed in the H$\alpha$ blue wing images, which may serve as indicators of localized energy release like Ellerman bombs. About 52\% of inside jets are associated with brightenings. However, no significant differences are found between associated and non-associated inside jets in terms of their physical properties. This may result from limited resolution or insufficient energy release for detection.

To assess the potential contribution of chromospheric jets to atmospheric heating, we estimate their typical thermal energy using \( E_{\text{thermal}} \sim \frac{3}{2} n k T V \), where \( n \) is the particle number density, \( k \) is the Boltzmann constant, \( T \) is the temperature, and \( V \) is the volume of the jet. Assuming a jet density of 10$^{15}$~cm$^{-3}$, a temperature of 5000~K, and a cylindrical shape with a length of 5~Mm and a width of 0.4~Mm, the order of the thermal energy of one event is 10$^{26}$~erg. As reported by \citet{2011ApJ...731...43N}, the estimated energy of chromospheric jets lies between those of nanoflares and microflares. Therefore, although chromospheric jets in active regions are unlikely to provide sufficient energy for chromospheric and coronal heating, small-scale energy release events occurring throughout the chromosphere likely play an important role in atmospheric heating.

The chromospheric jets studied in this work are driven by magnetic reconnection occurring in distinct magnetic field configurations. For inside jets, magnetic reconnection likely occurs between inclined magnetic fields within the penumbra. For outside jets, reconnection tends to take place between emerging flux and the ambient magnetic field. Following the classification by \citet{2015PhPl...22j1207S}, reconnection between emerging or moving magnetic fields, often referred to as anti-parallel reconnection, drives large-scale events such as chromospheric anemone jets and X-ray jets. In contrast, reconnection may occur between helical or twisted magnetic flux tubes oriented parallel or at an angle to each other. These configurations are considered more relevant to small-scale energy release events.

In this study, we investigate chromospheric jets located in and around the sunspot, which resemble previously reported classes of solar jets. The inside jets exhibit morphologies and characteristics similar to penumbral microjets, while the outside jets resemble chromospheric anemone jets. Compared to earlier studies that focused individually on penumbral microjets or chromospheric anemone jets, our work offers a unified observational framework for contrasting the two jet types within the same active region. Moreover, our investigation of penumbral microjets extends previous efforts by using H$\alpha$ blue wing images to examine the spatial correlation between jets and brightenings. A significant portion of inside jets is associated with brightenings, suggesting a potential link to localized energy release. While prior investigations found no significant response in coronal passbands associated with penumbral microjets \citep{2016ApJ...816...92T}, our detection of coronal brightenings in AIA images provides compelling evidence that magnetic reconnection may be responsible for driving these jets. Furthermore, our analysis of outside jets confirms established properties reported in existing literature on chromospheric anemone jets, yet the joint study of two types of chromospheric jets highlights their different physical properties and similar driving mechanisms. These findings contribute to a more comprehensive understanding of small-scale jet phenomena in active regions and offer new observational insights into the role of magnetic reconnection in the lower solar atmosphere.

\begin{acknowledgments}

The authors sincerely appreciate the data support provided by the NVST and SDO teams. This work is sponsored by the Strategic Priority Research Program of the Chinese Academy of Sciences, grant No. XDB0560000; the National Science Foundation of China (NSFC), under the numbers 12325303, 12473059, 12373115, 12203097, 11973084, and 12127901; the Yunnan Key Laboratory of Solar Physics and Space Science, under the number 202205AG070009; the Yunnan Fundamental Research Projects, under the numbers 202301AT070347 and 202301AT070349; the Yunnan Provincial Department of Education Science Research Fund Project, under the number 2025J0945; and the Chuxiong Normal University Doctoral Research Initiation Fund Project, under the number BSQD2420.

\end{acknowledgments}

\bibliography{References}{}
\bibliographystyle{aasjournalv7}

\end{document}